\documentclass[prl,twocolumn,showpacs,preprintnumbers,amsmath,amssymb]{revtex4-1}
\usepackage{graphicx}
\usepackage{epsfig}
\usepackage{dcolumn}
\usepackage{bm}

\usepackage{fix-cm}
\usepackage{epsfig,amsmath,amsfonts}
\newcommand{\Eq}[1]{Eq.~(\ref{#1})}
\newcommand{\Eqs}[1]{Eqs.~(\ref{#1})}
\newcommand{\eq}[1]{(\ref{#1})}

\begin{document}
\title{Viscosity of liquid $^4$He and quantum of circulation:
Why and how are they related? }

\author{ V. S. L'vov$^{1}$
      and L. Skrbek$^{2}$
          }
\affiliation{
$^1$Department of Chemical Physics, The Weizmann Institute of Science, Rehovot 76100, Israel}
\email{ victor.lvov@gmail.com}
\affiliation{
$^2$Faculty of Mathematics and Physics, Charles University,
Ke Karlovu 3, 12116 Prague, Czech Rep.}
\email{ skrbek@fzu.cz}
\date{\today}

\begin{abstract}
The relationship between the apparently unrelated physical quantities -- kinematic viscosity of liquid $^4$He, $\nu$, and quantum of circulation, $\kappa=2\pi \hbar/m_4$, where $\hbar$ is the Planck constant and $m_4$ denotes the mass of the $^4$He atom -- is examined in the vicinity of the superfluid transition occurring due to Bose-Einstein condensation. A model is developed, leading to the surprisingly simple relation  $\nu \approx \kappa/6$. We critically examine the available experimental data for $^4$He relevant to this simple relation and predict the kinematic viscosity for the stretched liquid  $^4$He along the $\lambda$-line at negative pressures.

\end{abstract}

\maketitle

In this Letter, following Lars Onsager \cite{Onsager} who was the first to raise this issue, we examine a  physical relation between two quantities of apparently completely different physical meaning: the kinematic viscosity of normal viscous fluids, $\nu$, responsible for their classical mechanical property -- friction between two fluid layers moving with different velocities; and the quantum of circulation,
\begin{equation}\label{kappa}
\kappa= 2\, \pi \hbar / M_{\rm s}\,,
\end{equation}
where $M_{\rm s}$ is the mass of the superfluid particle (such as the mass of $^4$He atom or two $^3$He atoms) describing quantization of velocity circulation in integer multiples of $\kappa$.
Both quantities $\nu$ and $\kappa$ have the same dimension, $[\nu]=[\kappa]=$~cm$^2$s$^{-1}$. This  rises the immediate question whether or not $\nu$ and $\kappa$ are related in fluids experiencing transition to superfluidity such as $^4$He (at $\lambda$-point $T=T_\lambda$),  $^3$He or Bose-Einstein condensates (BEC) of cold clouds of alkali atoms, and if so, why and  how? It should be noted that  $\nu$ is generally temperature and pressure dependent, while for a given medium $\kappa$ is a physical constant. We therefore bound the temperature region close to the physically relevant temperature $ T_*$ (which in the particular case of $^4$He is $T_*= T_\lambda$). Moreover, in order to guarantee the conventional viscous fluid behavior, we restrict the temperature region by $T>T_\lambda$ and, in order to neglect compressibility effects, we do not consider high external pressures far above the saturated vapor curve.

For \emph{classical} hydrodynamic community the question of a possible relationship between $\nu$ and $\kappa$ seems irrelevant.
Classical hydrodynamics does not involve the Planck constant, $\hbar$, and quantization of any relevant quantity does not take place.
At the same time, fluid viscosity, $\nu$, does not concern most researches in \emph{quantum} physics.
Therefore we should address this question to ``general" theoreticians and to the ``quantum" superfluid community.
In particular, this question is of high interest for researchers in the field of quantum turbulence \cite{Vinen,LS2010,VinenNew},
where the problem of turbulent energy dissipation in superfluids
is a hot topic \cite{Stalp02,Skr03,QFS05,TimPRL,ChGS,Golov1,Golov2,Bradley06,submittedJLTP,Kozik,Lvov1}.
It is usually discussed in terms of effective kinematic viscosity in units of $\gamma\kappa$,
where the prefactor $\gamma$ is expressed via phenomenologically introduced
parameters such as mutual friction coefficients or the Kolmogorov constant. At the same time, as far as we know, the relation between $\nu$ and $\kappa$ has not yet been seriously addressed.

\begin{figure}[t]
\begin{centering}
 \includegraphics[width=0.97\linewidth]{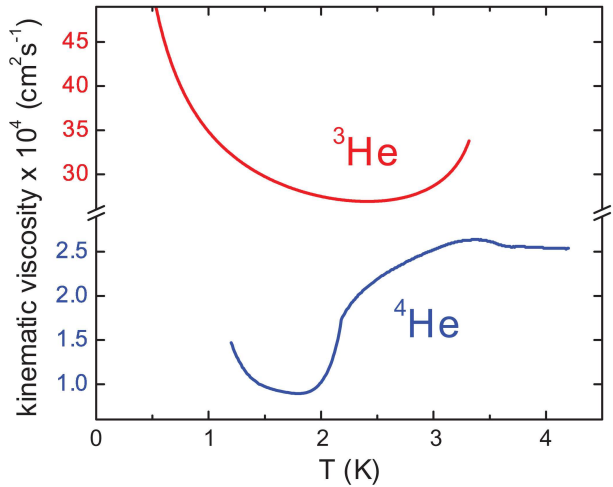}
\caption{(Color online) The temperature dependence of the kinematic viscosity  of liquid $^4$He (based on total fluid density) and liquid $^3$He (calculated using the dynamic viscosity as measured and fitted by Black, Hall and Thompson \cite{3HeVisc} and density values as analyzed and fitted by Huang, Chen, Li and Arp \cite{Huang}) along their SVP curves.  Note that the minimum $\nu$ value in normal liquid $^3$He exceeds that in normal liquid $^4$He by a order of magnitude.}
\label{nuT}
\end{centering}
\end{figure}

At  first glance, being of very different physical origin,  $\nu$ and $\kappa$  can be completely different. It is known (see Fig.~\ref{nuT}) from experiments that in the normal liquid He-I  $\nu$ first increases slightly along the saturation vapor pressure (SVP) curve from about  $2.59\times 10^{-4}$~cm$^2$s$^{-1}$ at its normal boiling point close to 4.2 K and then drops down to about  $1.67\times 10^{-4}$~cm$^2$s$^{-1}$ just above  $T_\lambda\approx 2.17$~K, i.e.,  $\nu$ varies by about 30\% and is of order $0.1 \,\kappa$ \cite{DB,Arp}. The superfluid community tends to consider this experimental fact as accidental~\cite{Vinen}, often having in mind the very different situation in fermionic $^3$He: Along the SVP the kinematic viscosity decreases slightly, but its lowest value is still about an order of magnitude bigger than that of He I in this range (see Fig.~\ref{nuT}), but with decreasing temperature $\nu$ increases and above the superfluid transition in the mK range it behaves as a thick Fermi liquid (for which the Landau Fermi liquid theory \cite{BP} predicts $\nu \propto T^{-2}$) with the (temperature and pressure dependent) kinematic viscosity comparable with that of air or machine oil, thus  $\nu / \kappa \gg 1 $ close to the critical temperature, $T_{\rm{c}}$.

``General" theoretical physicists think differently:  in the discussed temperature region both  $\nu$ and $\kappa$ (and $ T_\lambda$ as well) can  be linked uniquely using atomic units, such as the Planck constant $\hbar$, the atomic mass  $M$ or the Bohr radius. For $\nu$ and $\kappa$ this gives [in agreement  with \Eq{kappa}] $\nu\sim \kappa\sim \hbar/M$, or $\nu/\kappa \sim 1$, i.e. ``of the order of unity, at least  in Bose-fluids (such as $^4$He), while in Fermi-fluids (such as $^3$He) this is not necessarily the case~\cite{Grisha}".

This answer, for $^4$He in particular, cannot completely satisfy ``practically oriented researchers" for whom  numbers such as $4 \pi\approx 12$ or $1/{2\pi}\approx1/6$ (which can often be found in physical equations) cannot be considered as having the same order of magnitude, although  from the general theoretical viewpoint this is considered to be the case.

In this Letter, we use very simple physical arguments instead of just dimensional reasoning in order to find some more quantitative relationship between $\nu$ and  $\kappa$  rather then just $\nu \sim \kappa$, especially for the particular case of the most common quantum fluid - liquid $^4$He.

\textbf{The model.}
We start with the Boltzmann kinetic theory for weakly non-ideal gases, in frame of which $\nu$ can be estimated as:
\begin{subequations}\label{Nu}\begin{equation}\label{nu}
\nu \approx  \frac 13 \, v\, \lambda\ .
\end{equation}
Here $\lambda$ denotes the mean-free path and $v$ is the thermal velocity, at temperature $T \geq T_*$
given by
\begin{equation}\label{v}
\frac{M v^2}2 = \frac 32 \, k_{\rm{B}}T \,,
\end{equation}
where  $ k_{\rm{B}}$ is the Boltzmann constant and  $M$ denotes the atomic mass.
Equations~\eq{nu} and \eq{v} give
 \begin{equation}\label{nu1}
\nu  (T) \approx \sqrt{\frac{ k_{\rm{B}} T}{3\, M}}\, \lambda\ .
\end{equation}\end{subequations}

Let us consider liquid $^4$He and substitute for $T$ its particular value $T_* \cong T_\lambda$, estimated by the celebrated London-1938 formula for $T_{_{\rm  L}}$,  the temperature of Bose condensation of ideal gas ~\cite{T-L}:
 \begin{equation}\label{T-L}
 T_\lambda \approx T_{_{\rm  L}}=\left ( \frac{n}{\zeta(3/2)} \right )^{2/3} \,\frac{2\pi \hbar^2}{k_{\rm{B}} M} \cong 3.31\,  \frac{\hbar^2 \, n^{2/3} }{ k_{\rm{B}} M} \, ,
\end{equation}
where $\zeta(x)$ is the Riemann zeta function; $\zeta(3/2)\approx 2.6124$ and $n$ denotes the number of bosons in a unit volume, $n=N/V$.
Based on the experimental data available at that time, London   evaluated  from  his Eq. (\ref{T-L})  $T_{_{\rm  L}}\approx 3.1$~K. As it was pointed out many times, taking into account the simplicity of the ideal Bose gas model, this value is strikingly close to the experimentally observed $T_{\lambda}\approx 2.2 $~K. Having in mind that the temperature enters \Eq{nu1} under the square root, one  estimates the possible error of this replacement as $\sqrt{ T_{_{\rm  L}}/T_\lambda }-1\simeq 0.2$, i.e.,  within about 20\%. Optimists may say that this disagreement (up to 20\%),  originating from interatomic interactions, may also be considered as an inaccuracy estimate for our simple model of the kinematic viscosity.

 The next step is to estimate the mean-free path, $\lambda $, which in weakly interacting gas exceeds the mean interatomic distance $\ell  =n^{-1/3} $ considerably. Due to physical reasons (that lie outside the scope of this discussion), $\lambda \approx \ell$ in the densely packed liquid $^4$He at temperatures  $T \lesssim T_\lambda$, with a good accuracy (much better than 20\%). In this way, we implicitly account for strong interaction between liquid-helium atoms, thus exploiting the applicability limit of the Boltzmann Eq.~(\ref{nu}).

Now, \Eqs{nu1}, \eq{T-L} and \eq{kappa} allow one to estimate the kinematic viscosity of $^4$He just above its superfluid transition as
\begin{equation}\label{nu2}
 \nu_{\rm s}(T_\lambda)\approx \sqrt{\frac{3.31}{3}} \frac {\hbar }{M} = \sqrt{\frac{3.31}{3}} \,\,\frac \kappa {2\pi} \approx 0.167\,\, \kappa\approx \frac{\kappa}{6} \ .
 \end{equation}

Let us compare this simple model prediction with the available experimental data.
In fact, the thermodynamical and transport properties of cryogenic $^4$He are very well known. Along the saturated vapor pressure (SVP) curve (see Fig.~\ref{nuT}) they have been tabulated by Donnelly and Barenghi~\cite{DB}; we use their value of  $T_{\lambda}$ in Table~I.  Equilibrium thermodynamical properties of $^4$He can also be accurately calculated, thanks to Arp and McCarty, using the program HEPAK \cite{Arp}; which we have used to evaluate values of kinematic viscosity, $\nu$, at all positive pressures given in Table~I.

Having in mind the  approximative character of our reasoning, the accuracy of which we optimistically estimated as 20\%,  we have to  consider the surprisingly exact agreement  between the estimate~\eq{nu2} and the experimental data at saturated vapor pressure (SVP) given in Table I as unexpectedly good and accidental. In any case, even 20\% of inaccuracy is a  much better result than the simple dimensional estimate   $\nu(T_\lambda)\sim  \kappa$.

What can we say in brief about London's estimate of $T_\lambda$, used in our model? As we mentioned, he  found  $T_{\rm{L}}$ about 3.1~K in the approximation of ideal Bose gas which corresponds to the low density limit, when interatomic interactions can be neglected. On the other hand, it follows from the influential theoretical work of Bogolyubov on weakly non-ideal Bose gas \cite{Bog} that interactions are essential for the phenomenon of superfluidity, as for ideal Bose gas with quadratic dispersion relation the well-known Landau criterion \cite{Lan1941} gives zero critical velocity! From a practical point of view, the lowest possible density corresponds to metastable liquid along the nucleation line close to the negative spinodal pressure, at which  $T_\lambda\approx 2.2$ K, which is again  quite close to the London estimate.  The approximation of ideal gas gradually fails with increasing density.  It is therefore natural to think that the experimentally observed $T_{\lambda}$ will deviate from the predicted  $T_{_{\rm  L}}$ with increasing density of liquid $^4$He, i.e., with externally applied pressure, as the real system of liquid $^4$He deviates more and more from the ideal model of non-interacting Bose gas. This is indeed the case: the experimentally observed values are not proportional to the density in power 2/3 as predicted by London, moreover, with increasing pressure they  even decrease, making the slope $dp/dT_{\lambda}(p)$ negative. To illustrate this, we give the observed values of $T_{\lambda}$ at several elevated pressures, $p$, based on the accurate measurements of Ahlers \cite{Ahlers} in Table~I.

\begin{table}
 \begin{center}
\begin{tabular}{|c c c c c c|}
\hline
        &      &        &      &  & \\
~~$p$~~ & ~~$T_{\lambda}$~~   & $\rho$ & $\nu$ & $\nu/\kappa$  &\\
  bar        & K   &     g cm$^{-3}$     &      cm$^2$s$^{-1}$   &   & \\
\hline
25.868  &  ~~1.836~~    &  ~~0.1765~~  &  ~~$3.26\times10^{-4}$~~  &  ~~0.327~~   &\\
22.533  &  ~~1.889~~    &  ~~0.1737~~  &  ~~$3.04\times10^{-4}$~~  &  ~~0.305~~   & \\
18.180  &  ~~1.954~~    &  ~~0.1699~~  &  ~~$2.75\times10^{-4}$~~  &  ~~0.276~~   & \\
15.031 &  1.998      &  0.1669     &  $2.54\times10^{-4}$    &  0.255  & \\
7.328&  2.095      &  0.1580    & $2.02\times10^{-4}$   &   0.203  &\\
1.646&  2.157  &  0.1492  &  $1.72\times10^{-4}$  &  0.173 &   \\
SVP  &  2.172 &  0.1461  &      $1.67\times10^{-4}$ &  0.168 &\\
spinodal  &     $\simeq$~2.2   & $\simeq$~0.116   &  $\simeq~1.6\times10^{-4}$ & 0.167 &\\
\hline

\end{tabular}
\caption{ Selected properties of liquid  $^4$He along the $\lambda$-line. For details, see the text. }
\end{center}
\end{table}

\begin{figure}[t]
\begin{centering}
 \includegraphics[width=0.9\linewidth]{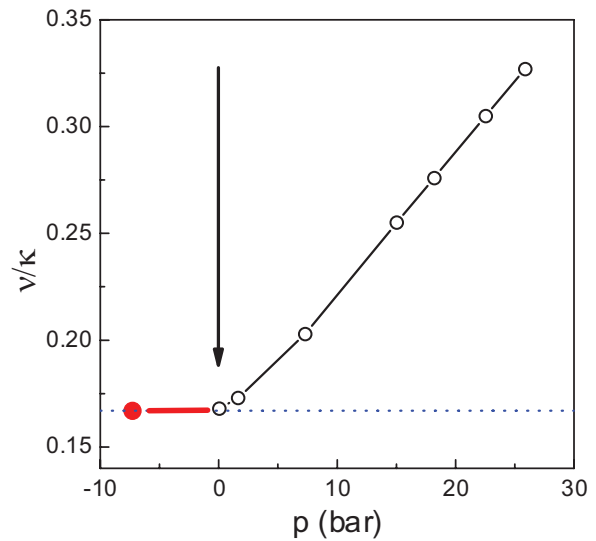}
\caption{(Color online) The ratio $\nu/\kappa$ plotted versus external pressure using values of Table I compared with the value predicted by Eq.~(\ref{nu2}) shown as dotted (blue) line. Negative pressures down to spinodal pressure, where metastable liquid could exist (red, left of the vertical arrow marking SVP), are included. Our approach allows to estimate the kinematic viscosity (see the red thick solid line) for the metastable liquid along the $\lambda$--line, which ought to extend down to the nucleation line. Its position in the pressure-temperature phase diagram was calculated \cite{ME}, but not as yet experimentally accurately determined.  }
\label{nukappa}
\end{centering}
\end{figure}

Fig.~\ref{nukappa} plots the dimensionless ratio $\nu/\kappa$ for liquid $^4$He at temperatures where the superfluid transition
is observed at various applied pressures. For the circulation quantum we use $\kappa= 0.997 \times 10^{-3}$ cm$^2$s$^{-1}$. We see that
$\nu/\kappa$ behaves in a regular way and with decreasing the external pressure approaches the value $\cong0.167$ given by Eq.~(\ref{nu2}).
It appears that our simple model works surprisingly well at low external pressures. This fact enables us to predict the value of kinematic viscosity for metastable liquid $^4$He that could exist at lower pressure than the equilibrium SVP and even at negative pressures down to the nucleation line, where cavitation occurs (for review on cavitation in liquid helium, see Balibar \cite{Balibar} and references therein). The location of the $\lambda$--line at negative pressures has been calculated by Maris and Edwards \cite{ME}. The maximum value of $T_{\lambda}\cong2.205$~K is reached at a negative pressure of about -5.7 bar, and the $\lambda$--line meets the spinodal line at a negative pressure of -7.3 bar and a temperature of about 2.18 K. Along this $\lambda$--line in the metastable liquid, the kinematic viscosity value ought to be close to $0.167\,\,\kappa$.

\textbf{Bosonic substances.}
Our approach would allow, in principle, the prediction of kinematic viscosity for other liquid bosonic substances in which the phenomenon of Bose-condensation could occur. However, clouds of spin polarized hydrogen or alkali atoms at conditions when BEC happens represent very dilute gases and our above estimate of mean free path cannot be used here.
For \textbf{fermionic substances} the situation is more complicated~\cite{fermi}.

We are fully aware that our approach could be rightfully criticized, both from the standpoint of general theoretical physics as well as from the point of view of classical fluid dynamics. Nevertheless, having in mind that theory of superfluid turbulence (especially problems of turbulent energy and/or vortex line density decay involving phenomenologically introduced quantity - effective kinematic viscosity) is still in its infancy, we hope that our approach can be considered as its useful building block.

To conclude, we have considered the physical relation between the two apparently independent physical quantities - kinematic viscosity, $\nu$,  versus quantum of circulation, $\kappa$ - and developed a simple model predicting that for bosonic liquids at the temperature of Bose condensation they are related as  $\nu \approx \kappa/6$. We have critically examined the experimental data on the transition to superfluidity at various external pressures for the most common bosonic liquid -- $^4$He -- and found that in the relevant region of
parameters ($T\simeq T_\lambda$ and at low applied pressures) the agreement is $\approx10^3$ times better then one may expect from dimensional analysis, which is usually accurate within about an order of magnitude. We also predict that this simple relation ought to hold even better for the metastable stretched liquid along the $\lambda$--line at negative pressures. One can estimate the probability of this agreement, $\approx10^{-3}$, as totally accidental. This raises many questions, for example:  (i)  Why does it happen?  (ii) Why do ideal-gas considerations work so well for the real liquid?  (iii) Could it be thanks to cancelation of effects of some as yet not known factors, which are definitely ignored in our suggested  approach? (iv) Are these cancelations accidental or are there some deep physical reasons yet to be discovered?  In other words, our Letter asks more general questions than it provides clear answers. This, we hope, will trigger new discoveries in fields of physics that are considered old and apparently well established.

We thank  R.J. Donnelly, W.F.~Vinen, D.~Schmoranzer, K.R.~Sreenivasan  G.E.~Volovik and two anonymous referees for stimulating discussions and constructive criticism. Partial support of the EU Research Infrastructures under the FP7 Capacities Specific Programme, MICROKELVIN  (project \# 228464) is highly appreciated. VSL acknowledges support of the U.S. - Israel BSF (grant \# 2008110); the research of LS is supported by the research plan MS 0021620834 of the Czech Republic and by GA\v{C}R 202/08/0276.


\begin{thebibliography}{99}
\bibitem{fermi} In case of fermionic substances we need to consider two characteristic temperatures of purely quantum mechanical origin: the Fermi temperature $T_{\rm F}$ and the critical temperature $T_{\rm c}$ at which the superfluid transition occurs (usually $T_{\rm c} \ll T_{\rm F}$). The crossover from classical behavior at high temperatures to quantum mechanical behavior at low temperatures is demarcated by $T_{\rm F}$, as the subsequent superfluid (superconducting) transition (in conventional BCS sense) is a purely quantum mechanical effect.
 Let us consider $^3$He and substitute for $T$ its particular value $T_* \cong T_{\rm F}$, estimated by the well known formula for the Fermi energy of the ideal Fermi gas:
$ E_{\rm F} = k_{\rm{B}}T_{\rm F}= \hbar^2\left (6\pi^2 n \right )^{2/3}/(2M)$ .
In analogy with the approach outlined above we formally arrive at
$ \nu(T_{\rm F})\approx  \kappa/ \left ({6 \pi^4} \right)^{1/6}  \approx 0.346 \,\, \kappa\,\,,$
where we took into account that the mass of the superfluid particle is $M_{\rm s} = 2 M$; for $^3$He $\kappa\approx0.66 \times 10^{-3}$~cm$^2$s$^{-1}$.
By comparison with a minimum in Fig.~\ref{nuT} we see that for the fermionic $^3$He the predicted value of kinematic viscosity is about an order of magnitude off. This disagreement is perhaps not too surprising, however, because superfluidity occurs at the transition temperature three orders of magnitude lower, where quantum mechanical effects are dominant and our starting assumption that the kinematic viscosity can be estimated from the Boltzmann  kinetic theory for ideal gases is not justified. Instead, as it has been shown by Landau in frame of his  Fermi liquid theory (see, e.g.,~\cite{BP}), in the temperature region of interest kinematic viscosity behaves as $\nu \propto T^{-2}$ and as the temperature approaches $T_{\rm c}$ our simple model is not applicable.
\end{thebibliography}
\end{document}